\documentclass[12pt]{article}
\usepackage{latexsym,epsfig,amssymb,amsmath,cite,verbatim,mcite,units}

\usepackage[vcentermath]{youngtab}          

\newcommand{\cM}{{\cal M}}
\newcommand{\cN}{{\cal N}}

\newcommand{\cL}{{\cal L}}
\newcommand{\cV}{{\cal V}}
\newcommand{\cK}{{\cal K}}
\newcommand{\cW}{{\cal W}}

\newcommand{\bP}{\mathbb P}
\newcommand{\bQ}{\mathbb Q}
\newcommand{\eins}{\mbox{$1 \hspace{-1.0mm} \text{l}$}}

\newenvironment{matr}[1]
{\left[ \begin{array}{{#1}}}{\end{array} \right]}

\def\bfone{\relax{\rm 1\kern-.35em 1}}

\newcommand{\be}{\begin{equation}}
\newcommand{\ee}{\end{equation}}
\newcommand{\ben}{\begin{displaymath}}
\newcommand{\een}{\end{displaymath}}
\newcommand{\bea}{\begin{eqnarray}}
\newcommand{\eea}{\end{eqnarray}}

\newcommand{\bean}{\begin{eqnarray*}}
\newcommand{\eean}{\end{eqnarray*}}

\newcommand{\fp}{f^{\text{\tiny{(\!+\!)}}}}
\newcommand{\fm}{f^{\text{\tiny{(--)}}}}
\newcommand{\fo}{f^{\text{\tiny{(1)}}}}
\newcommand{\ftw}{f^{\text{\tiny{(2)}}}}
\newcommand{\fth}{f^{\text{\tiny{(3)}}}}

\newcommand{\Fp}{F^{\text{\tiny{(\!+\!)}}}}
\newcommand{\Fm}{F^{\text{\tiny{(--)}}}}
\newcommand{\Fo}{F^{\text{\tiny{(1)}}}}
\newcommand{\Ftw}{F^{\text{\tiny{(2)}}}}

\makeatletter \@addtoreset{equation}{section} \makeatother

\begin{document}

\begin{titlepage}

\begin{flushright}
\small ~~
\end{flushright}

\bigskip

\begin{center}

\vskip 2cm

{\LARGE \bf Metastable supersymmetry breaking \\[2mm] in extended supergravity} \\[6mm]

{\bf Andrea Borghese and Diederik Roest}\\

\vskip 25pt

{\em Centre for Theoretical Physics,\\
University of Groningen, \\
Nijenborgh 4, 9747 AG Groningen, The Netherlands\\
{\small {\tt \{a.borghese, d.roest\}@rug.nl}}} \\

\vskip 0.8cm

\end{center}

\vskip 1cm

\begin{center} {\bf ABSTRACT}\\[3ex]

\begin{minipage}{13cm}
\small

We consider the stability of non-supersymmetric critical points of general $\cN 
= 4$ supergravities. A powerful method to analyse this issue based on the 
sGoldstino direction has been developed for minimal supergravity. We adapt this 
to the present case, and address the conceptually new features arising for 
extended supersymmetry. As an application, we investigate the stability when 
supersymmetry breaking proceeds via either the gravity or the matter sector. 
Finally, we outline the $\cN = 8$ case.

\end{minipage}

\end{center}


\vfill

\end{titlepage}


\section{Introduction}

The study of critical points of supergravity continues to play an important role in string model building, both from the cosmological as well as from the holographic point of view. In the former, De Sitter solutions are a first step towards modelling inflation, while in the latter the properties of Anti-de Sitter solutions are important for the dual field theory. 

Stability is clearly an important aspect in employing such solutions. Whereas supersymmetry preserving solutions are naturally stable, this is not at all clear for non-supersymmetric solutions. Indeed, due to the myriad of scalar fields in generic supergravity theories, one always faces the danger that at least one of these represents an instability, and hence renders the solution unstable. This is particularly worrisome for extended supergravity theories. Indeed, all known dS critical points of both $\cN = 8$ and $\cN = 4$ supergravity in $D=4$ are unstable \cite{HW84, Westra1, RR}. Up to very recently, the same appeared to hold for non-supersymmetric AdS solutions \cite{Warnerold1}. However, in \cite{Warnernew2} it was found that the $\cN = 8$ $SO(8)$ gauged supergravity in fact has a non-supersymmetric and nevertheless perturbatively stable critical point. In view of these developments, it appears interesting to derive general statements about the metastability of critical points in extended supergravity, and hence their usefulness in various aspects of string model building. This paper aims to make a first step towards this goal.

The route that we will take to this end involves a method that was developed in the context of minimal supergravity. In that context, it was realised that the sGoldstino offers an interesting window on the stability of non-supersymmetric critical points \cite{GR1}. As we will review in more detail later, the sGoldstino is a direction in scalar space that is singled out exactly by supersymmetry breaking, and hence exists for any non-supersymmetric solution. Restricting the mass matrix of all scalars to this direction, one can derive necessary conditions for stability. These will generically not be sufficient, as there can be tachyons in other scalar directions. Indeed, we will argue that the sGoldstino in a sense is rather far away from the onset of instabilities. Nevertheless, it is the only direction that one can study separately in a general fashion. 
Applications in the context of string model building and/or inflation were considered in \cite{Covi1}. We will adapt this method to extended supergravity theories, and analyse to what extent general constraints can be derived. 

A number of conceptually new features appear when applying this method to extended supergravity theories. The first was already encountered in $\cN = 2$ supergravity \cite{Louis}: rather than one, there are $\cN^2$ sGoldstini directions. We will argue that these generally split up in a number of gauge directions and a number of physical directions. Only the latter can be used to derive stability conditions. Furthermore, the sGoldstini can belong to different types of multiplets, corresponding to different types of supersymmetry breaking. In the case of $\cN = 4$ we will encounter supersymmetry breaking in the gravity sector and/or in the matter sector. Finally, one always has non-Abelian gauge groups subject to generalised Jacobi identities. These features were not present in previously considered cases, and therefore it is not clear whether they also allow for a sGoldstino analysis. This is what we will adress in this paper explicitly for $\cN = 4$, while we outline the procedure for $\cN = 8$.

As an aside, let us first clarify exactly we mean by stability, as this is not directly obvious in curved space-times.  In Minkowski, it corresponds to the requirement of $m^2 \geq 0$ for all fields, where $m^2$ is the coefficient of the quadratic term in the Lagrangian. From the field theory point of view, fields with $m^2 < 0$ represent tachyons. Analogously, such fields correspond to non-unitary irreps of the Poincar\'e isometry group. However, in curved space-times these requirements are somewhat different. The most famous example is the Breitenlohner-Freedman (BF) bound $m^2 \geq \tfrac34 V$ on scalar masses in Anti-de Sitter \cite{BF}, where $V$ is the cosmological constant, or the value of the scalar potential in the critical point. It is related to the AdS radius by $V= - 3/L^2$. The generalisation to fields with other spins and the opposite value of the cosmological constant has been investigated from both the field theory as well as the group theory point of view (see e.g.~\cite{Higuchi} and \cite{Nicolai}, respectively). The results naturally agree and can be expressed in $(m^2,V)$-diagrams such as figure 1 for gravitini and scalar fields, respectively, that will be relevant for what follows.

\begin{equation}
\begin{picture}(500,130)(-70,5)
\put(0,130){$V$}
\put(90,78.5){$m^2$}

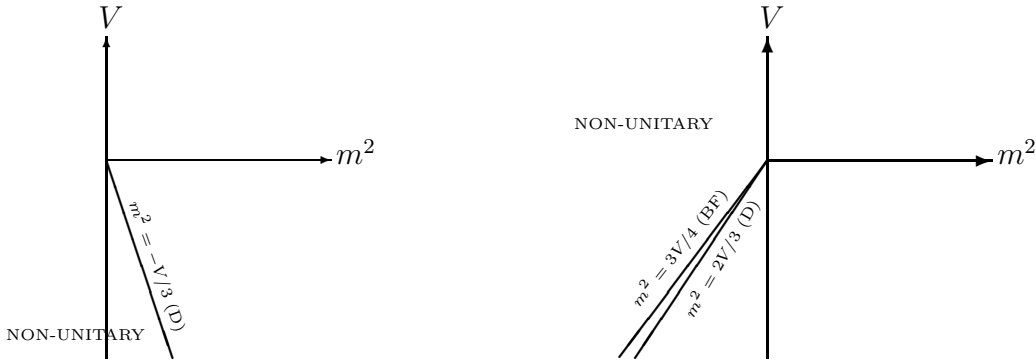
\put(-70,-15){{\small \it Figure 1: the $(m^2,V)$ diagrams of spin-3/2 and spin-0 fields, respectively. Adapted from \cite{Higuchi}.}}

\put(10,64){\rotatebox{290}{\tiny $ m^2=-V/3$ (D)}}
\put(-35,12){\tiny NON-UNITARY}



\put(3,80){\vector(1,0){85}}
\put(3,5){\vector(0,1){122}}

\thicklines

\put(3,80){\line(1,-3){25}}




\put(250,130){$V$}
\put(340,78.5){$m^2$}


\put(200,24){\rotatebox{52}{\tiny $ m^2=3V/4$ (BF)}}
\put(220,20){\rotatebox{60}{\tiny $ m^2=2V/3$ (D)}}


\put(180,92){\tiny NON-UNITARY}

\put(253,80){\vector(1,0){85}}
\put(253,5){\vector(0,1){122}}

\thicklines

\put(253,80){\line(-3,-4){56}}
\put(253,80){\line(-2,-3){50}}



\end{picture} \notag
\end{equation}

\vspace{0.8cm}

For spin-3/2 gravitini one finds that in De Sitter the bound is $m^2 \geq 0$. For all non-negative values the field has four propagating degrees of freedom. In both Minkowski and Anti-de Sitter, the bound is $m^2 \geq -V/3$. Above the bound the field again has four propagating degrees of freedom. Fields that saturate the bound, however, acquire an additional gauge invariance and only have two propagating degrees of freedom. In terms of group theory, this corresponds to a discrete unitary irreducible representation (UIR), while the right part of the diagram corresponds to a continuous family of UIRs.

For spin-0 scalars the bound in De Sitter coincides with that of Minkowski: $m^2 \geq 0$. In contrast, for AdS the masses can be negative. There is a discrete UIR giving rise to $m^2 = \tfrac23 V$, which is sometimes referred to as the conformal case. Note that this is above, rather than at, the BF bound of $\tfrac34 V$. In addition, there is a continuous family of UIRs with squared masses ranging from the BF bound to $+ \infty$. This phenomenon only takes place for scalars: for all fields with non-zero spin, the continuous family always has masses above a discrete UIR. We will see that both the UIRs at $m^2 = \tfrac34 V$ and $m^2 = \tfrac23 V$ play a special role in supergravity.

The outline of this paper is as follows. In section 2, we will review the sGoldstino approach for $\cN = 1$ supergravity. In section 3, we will outline the relevant features of $\cN =4$ supergravity, and discuss the dictionary between the two theories. Furthermore, we derive the full mass matrix of this theory. The stability of supersymmetric critical points is shown in section 4. Subsequently, section 5 addresses non-supersymmetric critical points. We derive the sGoldstino directions in general and derive the sGoldstini mass in two separate cases: those of supersymmetry breaking in the gravity and in the matter sector, respectively. We compare our results against the explicit examples that have appeared in the literature. Finally, we discuss our results and conclude in section 6. Our conventions and other useful expressions can be found in appendix A, while in appendix B we prove
the vanishing of the mass matrix when projected in the antisymmetric sGoldstini directions.

\section{Minimal supergravity}

It will be instructive to first discuss the sGoldstino approach within the framework in which it was originally developed, which is that of $\cN = 1$ supergravity. More details can be found in \cite{GR1}. 

Minimal supergravity in four dimensions allows for the following multiplets: a gravity multiplet plus a number of vector multiplets and a number of chiral multiplets. 
The scalars of the chiral multiplets span a K\"ahler space with metric $g_{i \bar \jmath} = \partial_i \partial_{\bar \jmath} \cK$, where $\cK$ is the K\"ahler potential. 
In the general case the possible deformations leading to a scalar potential are twofold. The first is characterized by a superpotential $\cW$, which is holomorphic in the chiral scalars, and leads to F-terms. The second are gaugings of the $U(1)$ R-symmetry (i.e.~Fayet-Iliopoulos terms \cite{VanProeyen}) and/or a number of isometries of the K\"ahler manifold, and leads to D-terms. The latter is only possible in the presence of vector multiplets.

We will first restrict ourselves to the case with a gravity multiplet coupled to a number of hypermultiplets. In this case the entire theory is characterized by the following combination of the K\"ahler and the superpotential: $\cL = e^{\cK/2} \cW$. 
The fermionic mass terms due to it are given by
 \begin{align}
   - \, \cL \, \psi^{\mu} \Gamma_{\mu \nu} \psi^{\nu} - \tfrac{i}{\sqrt{2}} \, \cL_{i} \, \chi^{i} \Gamma_{\mu} \bar \psi^{\mu} - \tfrac{1}{2} \, \cL_{ij} \, \chi^{i} \chi^{j} + \, {\rm h.c.} \, ,
 \end{align}
where $\cL_i = D_i \cL = \partial_i \cL + \partial_i \cK \, \cL$ is the $U(1)$-covariant holomorphic derivative. Higher-order covariant derivatives, such as $\cL_{ij}$, are defined in a similar way and are covariant with respect to the $U(1)$ R-symmetry and diffeomorphisms of the K\"ahler manifold. Furthermore, the scalar potential is given by
 \begin{align}
  V = - 3 \, |\cL|^2 + \cL_i \bar{\cL}^{i} \,.
  \label{N=1scalarpotential}
 \end{align}
The ensuing mass matrices for the chiral scalars are
 \begin{align}
\nonumber  & D_i D_{\bar \jmath} V = -2 \, g_{i \bar \jmath} |\cL|^{2} + \, \cL_{ik} \bar \cL_{\bar \jmath}{}^{k} - \, R_{i \bar \jmath k \bar l} \, \bar \cL^{k} \cL^{\bar l} + \, g_{i \bar \jmath} \, \bar \cL^{k} \cL_{k} - \, \cL_i \bar \cL_{\bar \jmath} \, ,\\
& D_i D_j V = - \, \cL_{ij} \bar \cL + \, \bar \cL^{k} \cL_{(ij)k} \, ,
 \label{N=1massmatrix}
 \end{align}
in terms of $\cL$, its covariant derivatives, and $R_{i \bar \jmath k \bar l}$ being the Riemann tensor of the K\"ahler manifold spanned by the chiral scalars. 

The different tensors appearing in these mass matrices play the following roles:
 \begin{itemize}
  \item
  $\cL$ - the scale of supersymmetric AdS: it lowers the value of the scalar potential and gives rise to a mass term of the gravitino. With regards to critical points of the scalar potential, this term does not break the supersymmetry of the corresponding AdS solutions.
 \item
 $\cL_i$  - the order parameter of supersymmetry breaking: it raises the value of the scalar potential and  gives rise to a term bilinear in the gravitino and the dilatini in the Lagrangian. In contrast to $\cL$, this term does break supersymmetry of the critical points of the scalar potential. The effect of supersymmetry breaking is to raise the value of the scalar potential, as can be seen from \eqref{N=1scalarpotential}.
 \item
 $\cL_{ij}$ - the supersymmetric mass term: it gives rise to the mass term of the dilatini and the chiral scalars; in other words, of all fields outside the gravity multiplet. It does not break supersymmetry of the critical points.
 \item
 $\cL_{ijk}$ - this term only appears in the mass matrix \eqref{N=1massmatrix}. It will drop out of what follows and hence will be irrelevant for the present discussion.
 \end{itemize}
Based on these interpretations, the critical points of the scalar potential divide into two classes: supersymmetric ones for which $\cL_i$ vanishes, and non-supersymmetric ones for which it does not.

The stability of the supersymmetric critical points is easiest to discuss. An arbitrary direction in scalar space, characterized by arbitrary vector $v^I = (v^i, \bar w^{\bar \imath})$ with independent vectors $v^i$ and $w^{i}$, reads in this case
 \begin{align}
  v^I m^2_{IJ} v^J = - \tfrac94 \, v^i \bar v_i |\cL|^2 + \tfrac14 \, (2 \, v^i \cL_{ik} -    {\bar w}_k \cL )(2 \, \bar v^{\bar \jmath} {\bar \cL}_{\bar \jmath \bar k} - w_{\bar k}  \bar \cL) + (v \leftrightarrow \bar w) \,.
 \end{align}
The first contribution is negative definite and gives rise to scalar masses exactly at the Breitenlohner-Freedman bound: $m^2  = \tfrac34 V$. The second contribution is positive definite. Therefore, and not surprisingly, we find that in this case all scalars are stable due to supersymmetry. Note that in the absence of the supersymmetric masses $\cL_{ij}$, the second term yields a contribution such that the total mass is $m^2 = \tfrac23 V$. As we have seen in the introduction and is illustrated in figure 1, this corresponds exactly to the discrete scalar representation of $SO(2,3)$. The introduction of $\cL_{ij}$ serves to lift the degeneracy of masses, and redistributes these to different values $m^2 \geq \tfrac34 V$.

The discussion of the stability of the non-supersymmetric critical point is somewhat less straightforward. Due to the complexity of the mass matrices with $\cL_i$ reinstated, it is very difficult to make general statements about the stability of all scalars. However, it is possible to focus on a particular scalar. This possibility arises as the very fact of breaking supersymmetry singles out a particular direction of the scalar manifold, being the sGoldstino. The argument is as follows. When breaking supersymmetry, the gravitino aquires an additional mass term and hence moves away from the line at $m^2 = -V/3$ in figure 1. While going from the discrete to a continuous irrep, it loses gauge invariance. The additional degrees of freedom are provided by a particular linear combination of the dilatini: by a slight abuse of notation, this is called the (would-be) Goldstino. This is the fermionic counterpart of the Higgs mechanism. The supersymmetric partner of the Goldstino is referred to as the sGoldstino. It is a particular linear combination of the scalar fields determined by the order parameter of supersymmetry breaking, which is $\cL_i$ in $\cN =1$. 

The sGoldstino mass is therefore given by the projection of the mass matrix \eqref{N=1massmatrix} with the particular vector $v^I = (0, \cL^{\bar \imath})$, which reads
 \begin{align}
   m^2 = +2 \, |\cL|^2 - \, R_{i \bar \jmath k \bar l} \, \bar{\cL}^i \cL^{\bar \jmath} \bar{\cL}^k \cL^{\bar l} \,.  \label{N=1-sG}
 \end{align}
A number of points are noteworthy. Firstly, this expression only depends on $\cL$, the scale of supersymmetric AdS, and the sectional curvature of the K\"ahler manifold in the direction $\cL_i$. Perhaps somewhat surprisingly at first sight, the sGoldstino mass does not approach $\tfrac23 V$ in the limit $\cL_i \rightarrow 0$. This comes about due to the extremality condition for the scalars,
 \begin{align}
  \cL_{ij} \bar{\cL}^{j} = 2 \, \cL_i \bar \cL \,.
 \end{align}
It relates the supersymmetric mass scale to the supersymmetric AdS scale. Therefore it is inconsistent to have only $\cL$ and its first derivative non-vanishing. The introduction of $\cL_{ij}$ raises the sGoldstino mass to $+ 2 |\cL|^2$ in the limit when supersymmetry is restored. Other components of $\cL_{ij}$, transverse to $\bar{\cL}^j$, will affect the masses of the complementary scalars but not of the sGoldstino.  Secondly, the third derivative term $\cL_{ijk}$ has also dropped out. 

One can subsequently analyse in which cases the sGoldstino mass is positive (or above the BF bound in AdS). This is only a necessary but not sufficient condition:  if the sGoldstino mass is negative (or below the BF bound in AdS) one has proven the instability of the critical point. If it is not, any of the complementary scalars could still be unstable. In particular, before the breaking of supersymmetry, the sGoldstino has a mass of $+ 2 |\cL|^2$ in the AdS case. It is therefore certainly not close to the BF threshold of instability. Other scalars might be closer and could therefore be more sensitive to supersymmetry breaking effects. The problem is that these complementary scalars cannot be addressed in a general way similar to the sGoldstino.

A possible approach towards the stability of Minkowski or De Sitter critical points based on \eqref{N=1-sG} is to divide $\cN = 1$ theories based solely on the sectional curvatures (and independently of the superpotential). If the sectional curvature in all directions is such that the sGoldstino mass can never be positive, irrespective of the superpotential and hence $\cL_i$, one has proven that all non-supersymmetric points are unstable. Note that this approach can not rule out non-supersymmetric yet metastable Anti-de Sitter critical points, as for such cases the positive contribution due to $\cL$ can always overcome any negative contributions due to the sectional curvature. This finding seems to resonate with the non-supersymmetric and stable critical point of $\cN = 8$ supergravity \cite{Warnernew2}.

The inclusion of vector multiplets leads to the additional possibility to turn on (positive definite) D-terms in the scalar potential. In this case, the would-be Goldstino is a linear combination of spin-1/2 fields of both the chiral and the vector multiplets. However, as the latter have no scalars as supersymmetric partners, the projection onto the sGoldstino scalars remains given by $\cL_i$. Performing this projection on the mass matrix in the presence of D-terms, one finds a more complicated expression which can be found in \cite{GR1}. For the present purpose it suffices to say that in such a case stability is easier to attain, as the D-terms raises the mass, but general statements are harder to derive.

\section{Half-maximal supergravity}

\subsection{Covariant formulation}

Half-maximal supergravity in four dimensions allows for the following multiplets: the gravity multiplet, and a number $n$ of vector multiplets. In contrast to the minimal theory, the scalar manifold is completely determined by supersymmetry and is given by the coset space
 \begin{align}
   \frac{SL(2)}{SO(2)} \times \frac{SO(6,n)}{SO(6) \times SO(n)} \,,
 \end{align}
The numerator of this expression is the global symmetry group of the theory. Due to the fact that the scalar manifold and hence its sectional curvatures are completely fixed for $\cN \geq 3$ supergravities, one would expect an analysis based on the analogon of \eqref{N=1-sG} to be very powerful.

In contrast to the minimal theory discussed in the previous section, $\cN = 4$ does not allow for the introduction of the analogon of an arbitrary superpotential. Due to extended supersymmetry, all possible deformations are induced by gaugings, and the corresponding deformations are determined by constant parameters: the embedding tensor \cite{Samtleben}. For $\cN = 4$ these are given by the following $SL(2) \times SO(6,n)$ tensors \cite{SW}:
 \begin{align}
 f_{\alpha\,  MNP} \,, \quad \xi_{\alpha \, M} \,, \label{emb-tensor}
\end{align}
where $\alpha$ and $M$ are $SL(2)$ and $SO(6,n)$ indices, respectively.
The former gauges a subgroup of $SO(6,n)$, while the latter always induces a gauging of $SL(2)$ as well. We will argue in the concluding section that the three-form $f$ should be thought of as F-terms, while the fundamental irrep $\xi$ is the $\cN = 4$ equivalent of D-terms and Fayet-Iliopoulos terms.

The introduction of these components has the following consequences for the Lagrangian. Firstly, all derivatives are covariantised with respect to the gauge group induced by the embedding tensor. Furthermore, fermion bilinear terms of the form (focussing on the gravitini and omitting the spin-1/2 bilinears)
  \begin{align}
  \tfrac13 g \, A_1^{ij} \, \bar \psi_{\mu i} \Gamma^{\mu \nu} \psi_{\nu j} - \tfrac13 i g \, A_2^{ij} \, \bar \psi_{\mu i} \Gamma^{\mu} \chi_j - ig \, \bar A_{2 \, a}{}^{i}{}_{j} \, \bar \psi_{\mu i} \Gamma^{\mu} \lambda^{a j} + \, h.c.
 \label{N=4-fermionbil}
 \end{align}
have to be included, where the tensors $A_{1,2}$ are given by
 \begin{align}
    A_1^{ij} &= \epsilon^{\alpha \beta} (\cV_{\alpha})^* \, \cV_{[kl]}{}^{M} \cV_{N}{}^{[ik]} \cV_{P}{}^{[jl]} \, f_{\beta \, M}{}^{NP} \, ,
\nonumber \\
   A_2^{ij} &= \epsilon^{\alpha \beta} \cV_{\alpha} \, \cV_{[kl]}{}^{M} \cV_{N}{}^{[ik]} \cV_{P}{}^{[jl]} \, f_{\beta \, M}{}^{NP} + \tfrac32 \, \epsilon^{\alpha \beta} \cV_{\alpha} \, \cV_{M}{}^{[ij]} \, \xi_{\beta}{}^{M} \, ,
\nonumber \\
   {A_{2\, a\, i}}^j &= \epsilon^{\alpha \beta} \cV_{\alpha} \, \cV_{a}{}^{M} \cV^{N}{}_{[ik]} \cV_{P}{}^{[jk]} \, f_{\beta \, MN}{}^{P} - \tfrac14 \,\epsilon^{\alpha \beta} \cV_{\alpha} \, \delta_{i}^{j} \cV_{a}{}^{M} \, \xi_{\beta \, M} \, .
 \end{align}
Finally, a scalar potential that is bilinear in the embedding tensor appears. In terms of $SL(2) \times SO(6,n)$ covariant quantities it reads
 \begin{align}
V = \tfrac{1}{16} \, \big\{ & f_{\alpha \, MNP} f_{\beta \, QRS} \, \cM^{\alpha \beta} \, \big[ \tfrac13 \, \cM^{MQ} \cM^{NR} \cM^{PS} + \big( \tfrac23 \, \eta^{MQ} - \cM^{MQ} \big) \, \eta^{NR} \eta^{PS} \big] + \nonumber \\
 & - \tfrac94 \, f_{\alpha \, MNP} f_{\beta \, QRS} \, \epsilon^{\alpha \beta} \cM^{MNPQRS} + 3 \, \xi_{\alpha}{}^{M} \xi_{\beta}{}^{N} \, \cM^{\alpha \beta} \, \cM_{MN} \big\} \,, \label{SW-scalarpot}
 \end{align}
where our conventions for the scalars are given in appendix A.

An essential role in the present discussion will be played by the quadratic constraints. These are bilinear conditions on the structure constants \eqref{emb-tensor}. The full set of quadratic constraints can be found in  \cite{SW}. For future purposes we list here the non-trivial ones after setting $\xi$ to zero. In terms of $SL(2) \times SO(6,n)$ covariant quantities they are given by
 \begin{align}
 f_{\alpha \, R[MN} f_{\beta \, PQ]}{}^{R} & = 0 \, , \quad
 \epsilon_{\alpha \beta} f_{\alpha \, RMN} f_{\beta \, PQ}{}^{R} = 0 \, . \label{QC}
 \end{align}
These ensure the consistency of the gauging. 

\subsection{Formulation in the origin}

We now use the following crucial property, in which $\cN =4$ supergravity differs from $\cN \leq 2$. Instead of retaining covariance with respect to the full symmetry group, we will use the non-compact generators in order to go to the origin in moduli space. This does not constitute a loss of generality as the moduli space is homogeneous. The remaining symmetry group is then the isotropy group
 \begin{align}
  SO(2) \times SO(6) \times SO(n) \,. \notag
 \end{align} 
We will use the indices $\alpha$, $m$ and $a$ for the different factors, respectively. Moreover, in the rest of this section, we will set $\xi$ equal to zero. The generalisation to non-zero $\xi$ is discussed in the conclusions.

The embedding tensor splits up in the following irreps of the reduced symmetry group, playing the following roles (as should become clear in what follows):
 \begin{itemize}
  \item
The scale of supersymmetric AdS is set by
 \begin{align}
  \fp \equiv \tfrac12 \, (f_{\alpha \, mnp} + \tfrac{1}{3!} \, \epsilon_{\alpha \beta} \epsilon_{mnpqrs} \, f_{\beta \, qrs}) \,.
 \end{align}
Note that this combination corresponds to the imaginary self-dual (ISD) irrep of $SO(2) \times SO(6)$ (similar to that appearing in $\cN = 1$ flux compactifications \cite{GKP}). This combination shows up in the $SU(4)$ matrix $A^{ij}_1$, and hence also gives rise to the gravitini masses via the $SU(4)$ matrix. In this way, upon turning on $\fp$, all gravitini of the theory move from the origin of figure 1 to a point on the $m^2 = -V/3$ line, corresponding to supersymmetric Minkowski and AdS, respectively.
 \item 
The order parameter of supersymmetry breaking in the gravity sector is given by
  \begin{align}
  \fm \equiv \tfrac12 \, (f_{\alpha \, mnp} - \tfrac{1}{3!} \, \epsilon_{\alpha \beta} \epsilon_{mnpqrs} \, f_{\beta \, qrs}) \,.
 \end{align}
Turning on the anti-imaginary self-dual (AISD) component $\fm$ implies a non-zero $SU(4)$ matrix $A^{ij}_2$. For that reason, the gravitini acquire an additional mass term by absorbing the spin-1/2 fields of the gravity multiplet.
 \item
The order parameter of supersymmetry breaking in the matter sector is
 \begin{align}
  \fo \equiv f_{\alpha \, mna} \,.
\end{align} 
The component $\fo$ induces a non-zero $SU(4)$ matrix ${A_{2\, a\, i}}^j$. In this case, additional mass terms for the gravitini arise due to the coupling to spin-1/2 fields in the vector multiplets.
 \item
The supersymmetric mass terms are
 \begin{align}
  \ftw \equiv f_{\alpha \, mab}
 \end{align} 
This component induces mass terms for the matter sector (both the spin-1/2 fields and the scalars of the vector multiplets).
 \item
Finally, we have
  \begin{align}
   \fth \equiv f_{\alpha \, abc} \,.
  \end{align}
In analogy with $\cL_{ijk}$ in the $\cN =1$ case, this component only appears in the mass of the matter scalars multiplied by $\fo$.  Indeed we will see that the sGoldstini mass can always be written in such a way that it is independent of this component.
 \end{itemize}
In addition to these five tensors, we will denote their bilinear contractions by $F$-tensors. Due to the (A)ISD properties of $\fp$ and $\fm$, their bilinears will also satisfy a number of non-trivial properties. More details on our conventions can be found in appendix A.

The scalar potential in the origin reads
 \begin{align}
  V & = - \tfrac14 \, \Fp + \tfrac{1}{12} \, \Fm + \tfrac14 \, \Fo \,,
 \end{align}
and hence is completely determined by the scale of SUSY AdS plus SUSY breaking effects, both from the gravity and the matter sector.

The vanishing of the first derivatives leads to the extremality conditions
 \begin{align}
 \Fo_{\alpha, \, \beta} - \tfrac12 \, \delta_{\alpha \beta} \Fo & = \tfrac23 \, F^{\text{\tiny{(\!+ --)}}}_{(\alpha, \, \beta)} \, , \qquad
F^{\text{\tiny{(1 2)}}}_{m,\,a} = F^{\text{\tiny{(\!+\! 1)}}}_{m,\,a} - F^{\text{\tiny{(-- 1)}}}_{m,\,a} \, . \label{N=4-fieldeqs}
 \end{align}
The former equation is symmetric and traceless, and follows from the $SL(2)$ scalars, while the second equation corresponds to the non-compact scalars of $SO(6,n)$. Also note that both these equations are automatically satisified in the SUSY case with $A_2 = 0$.

Turning to the second derivative of the scalar potential, we find the following results
\begin{align}
V_{\alpha \beta, \, \gamma \delta} = & (\delta_{\alpha \gamma} \delta_{\beta \delta} + \delta_{\alpha \delta} \delta_{\beta \gamma} - \delta_{\alpha \beta} \delta_{\gamma \delta}) \, (- \tfrac{1}{12} \, \Fp - \tfrac{1}{12} \, \Fm + \tfrac14 \, \Fo) \,, \label{SL2-mass-1} \\
V_{\alpha \beta, \, bn} = & \tfrac12 \, \epsilon_{\gamma (\alpha} \, ( F^{\text{\tiny(1 2)}}_{\gamma n, \, |\beta) b} + F^{\text{\tiny(1 2)}}_{\beta) n, \, \gamma b}) \,, \label{mixed-mass-1} \\
V_{am, \, bn} = & \tfrac14 \, (- \, \delta_{ab} \, \Fp_{m,\,n} + \, \delta_{ab} \, \Fm_{m,\,n} + \, \delta_{ab} \, \Fo_{m,\,n} + \, \delta_{mn} \, \Ftw_{a,\,b} - \, \Fo_{an, \, bm} - \, \Ftw_{an, \, bm} + \nonumber \\ 
& \quad - \, F^{\text{\tiny{(\!+\! 2)}}}_{mn,\,ab} + 3 \, F^{\text{\tiny{(-- 2)}}}_{mn,\,ab} + \, F^{\text{\tiny(1 3)}}_{mn,\,ab}  - \tfrac12 \, \epsilon_{\alpha \beta} \epsilon_{mn p_1p_2p_3p_4} \; f_{\alpha \, ap_1p_2} f_{\beta \, bp_3p_4}) \,. \label{SO66-mass-1}
\end{align}
Using the relations in appendix A, we can turn to physical fields $\phi^{I} = \{\chi, \phi, \phi^{\{am\}}\}$
\begin{align}
& V_{\chi,\, \chi} = V_{\phi,\, \phi} = - \tfrac{1}{12} \, \Fp - \tfrac{1}{12} \, \Fm + \tfrac14 \, \Fo \,, \label{SL2-mass-2} \\
& V_{\chi, \, \{bn\}} = \sigma_{3}^{\alpha \beta} \, V_{\alpha \beta,\, bn} \, , \qquad V_{\phi, \, \{bn\}} = \sigma_{1}^{\alpha \beta} \, V_{\alpha \beta,\, bn} \, , \label{mixed-mass-2} \\
& V_{\{am\},\,\{bn\}} = 4 \, V_{am , \, bn} \, , \label{SO66-mass-2}
\end{align}
where $\sigma_{1}, \sigma_{3}$ are the standard Pauli matrices.
In order to compute the squared mass matrix, we have to multiply the Hessian matrix by the inverse K\"ahler metric
\begin{align}
[ m^{2} ]_{I}{}^{J} = \frac{\partial^{2} V}{\partial \phi^{I} \partial \phi^{K}} \, [\cK^{-1}]^{KJ} \, .
\end{align}
The K\"ahler metric can be read explicitly from the kinetic terms in appendix A and, in our case, is given by
\begin{align}
\cK_{IJ} = \begin{matr}{cc} \tfrac12 \eins_{2} & 0 \\ 0 & \eins_{6n} \end{matr} \, .
\end{align}

Despite these somewhat complicated expressions, in the following we will derive a number of non-trivial bounds from the mass matrices. In particular, we will analyse the following cases. To the best of our knowledge, all known critical points in the literature either have all three tensors $\fp, \fm, \fo$ non-vanishing (see e.g.~\cite{RR}), or only one of these non-vanishing (see e.g.~\cite{Westra1}). The latter case has either full supersymmetry with $\fp$, or has fully broken supersymmetry in the gravity sector due to $\fm$ or the matter sector due to $\fo$. We will focus on these three cases in this paper, and leave the analysis with all three tensors non-vanishing for future work. For this reason, we will not consider partial supersymmetry breaking, as recently discussed for $\cN =2$ in \cite{Smyth}.

\section{Supersymmetric critical points}

As a consistency check, we will first discuss the stability of supersymmetric critical points. This analysis will follow its $\cN = 1$ counterpart very closely. Setting $A_2$ and hence $\fm$ and $\fo$ equal to zero, the mass matrices are as follows. 
\begin{align}
& \left[m^{2}\right]_{\chi}{}^{\chi} = \left[m^{2}\right]_{\phi}{}^{\phi} = - \tfrac{1}{6} \, \Fp \,, \nonumber \\
& \left[m^{2}\right]_{\{am\}}^{\quad \; \{bn\}} =  - \delta_{ab} \, \Fp_{m,\,n} + \, \delta_{mn} \, \Ftw_{a,\,b} - \, \Ftw_{an, \, bm} - \, F^{\text{\tiny{(\!+\! 2)}}}_{mn,\,ab} \,,
\end{align}
while the scalar potential is given by $- \tfrac14 \Fp$.

For the $SL(2)$ case we find that the masses are exactly equal to $\tfrac23 V$. As was discussed in the introduction, this value corresponds to the discrete irrep, and does not saturate the BF bound. Note that for the $SL(2)$ scalars, there are no SUSY mass terms to lift the degeneracy. This is a consequence of being in the gravity multiplet: in the SUSY case all fields in this multiplet correspond to discrete irreps.

For the $SO(6,n)$ scalars the mass matrix is somewhat more interesting.
Again, in the absence of SUSY mass terms, the masses are given by $m^2 = \tfrac23 V$. Interestingly, including $\ftw$ as well, the mass matrix can be rewritten in the following form:
 \begin{align}
\left[m^{2}\right]_{\{am\}}^{\quad \; \{bn\}} = - \tfrac{3}{16} \, \delta_{ab} \delta_{mn} \, \Fp + \tfrac18 \, V_{am, \, \alpha \, pqc} V_{bn, \, \alpha \, pqc} \,,
 \end{align}
where
 \begin{align}
   V_{am, \, \alpha \, pqc} = - 4 \delta_{m[p} f_{\alpha \, a|q]c} + \delta_{ac} \fp_{\alpha \,  mpq} \,.
 \end{align}
The latter term is clearly positive definite. Therefore the former term in the mass matrix sets a lower bound for the masses at $m^2 = \tfrac34 V$, which is at (rather than above) the BF bound. Turning on the SUSY mass terms therefore again serves to break the degeneracy between the different $SO(6,n)$ masses, and indeed could lower the mass up to the BF bound. To saturate rather than satisfy the bound for some scalar field one needs to have a direction $U^{am}$ such that it annihilates the additional term; in other words one should have $U^{am} V_{am, \, \alpha \, pqc}$ to be vanishing. Note that this story is completely analogous to the $\cN = 1$ case - there one also finds $\tfrac23 V$ for the masses if one only turns on $\cL$, but turning on SUSY mass terms $\cL_{ij}$ this can change into $m^2 \geq \tfrac34 V$.

\section{Non-supersymmetric critical points}

\subsection{sGoldstini directions}

Now let us turn to non-supersymmetric points. In analogy to the $\cN =1$ case, one can read off the Goldstini directions from the fermion bilinear terms in \eqref{N=4-fermionbil}. They are given by
 \begin{align}
    \tfrac{i}{6} \, A_2^{ij} \, \chi_j + \tfrac{i}{2} \, \bar{A}_{2a}{}^i{}_j \, \lambda^{aj} \,,
 \end{align}
where the spin-1/2 shift matrices in the origin of the coset space read 
 \begin{align}
   A_2^{ij} = \epsilon_{\alpha \beta} \cV_{\alpha} \, f_{\beta \, [kl]}{}^{[ik][jl]} \, , \qquad    \bar A_{2\, a}{}^{i}{}_{j} &= \epsilon_{\alpha \beta} (\cV_\alpha)^* \, f_{\beta \, a}{}^{[ik]}{}_{[jk]} \, .
 \end{align}
The sGoldstini correspond to the supersymmetric scalar partners of the Goldstini. These amount to
 \begin{align}
  \tfrac{1}{6} \, A_2^{ij} \tau - \tfrac{1}{2} \, \bar{A}_{2a}{}^i{}_k \, [G_{m}]^{kj} \, \phi^{\{am\}} \, ,
 \end{align}
where the field $\tau$ is a complex combination of $\chi$ and $\phi$, and we use the 't Hooft symbols as given in the appendix A to go from the $\bf 6$ representation of $SU(4)$ to that of $SO(6)$. In terms of the latter, the 16 sGoldstini directions are the following:
 \begin{align}
  V^{ij}_{\tau} & = \, \tfrac{1}{48} \,\epsilon_{\gamma \eta} \cV_{\gamma} \, \fm_{\eta \, mnp} \, [G_{mnp}]^{ij} - \tfrac{1}{8} \, \epsilon_{\gamma \eta} \cV_{\gamma} \, \xi_{\eta \, m} \, [G_{m}]^{ij} \,, \notag \\
  V^{ij}_{\{am\}} & = - \tfrac{1}{8} \, \epsilon_{\gamma \eta} (\cV_{\gamma})^* \, \fo_{\eta \, anp} \, [G_{mnp} - 2 \delta_{m[n} G_{p]}]^{ij} + \tfrac{1}{8} \, \epsilon_{\gamma \eta} (\cV_{\gamma})^* \, \xi_{\eta \, a} \, [G_{m}]^{ij} \,. \label{sGolds}
 \end{align}
Note that, in the general case, the Goldstini and sGoldstini comprise fields from both the gravity and the vector multiplets. This corresponds to the new feature of $\cN =4$ supergravity to have supersymmetry breaking in both the gravity and matter sectors, for $\fm$ and $\fo$ non-vanishing, respectively.

The product of the two fundamental $SU(4)$ representations $i$ and $j$ splits up in two irreps, corresponding to the symmetric and anti-symmetric parts. In terms of 't~Hooft symbols, these correspond to the $G^{(3)}$ and $G^{(1)}$ terms, respectively. The following interpretations seem to hold for these two irreps:
 \begin{itemize}
  \item
The six sGoldstini directions given by the anti-symmetric combination are to be interpreted as gauge transformations. That is, these directions in scalar space have been gauged away by the introduction of the associated embedding tensor components. The associated gauge vectors are those of the gravity multiplet. This can be seen in two ways. Firstly, the form of the antisymmetric part of \eqref{sGolds} coincides with an explicit gauge transformation on the scalars in the origin. Secondly, it can be checked that the scalar mass matrix indeed is annihiliated up by these directions:
 \begin{align}
  m^2 \cdot V^{[ij]} = 0 \,, \label{sG-gaugeinv}
 \end{align}
where $m^2$ is the mass matrix corresponding to both gravity and matter scalars. The explicit proof of this can be found in appendix B.
\item
The ten scalar directions that are symmetric constitute the physical sGoldstini directions. These can be used to infer statements about stability from the mass matrices. Instead of considering the eigenvalues of all ten directions, we will focus on the only mass condition that is $SU(4)$ invariant. This corresponds to taking the trace over all sGoldstini:
 \begin{align}
  M^2_{\rm sG} \equiv \bar{V}_{(ij)} \cdot m^2 \cdot V^{(ij)} \,. \label{sG-symm}
 \end{align}
\end{itemize}
We expect the interpretation of the symmetric and anti-symmetric sGoldstini as physical and gauge directions, respectively, to hold for other supergravity theories as well.  In terms of Young tableaux of the R-symmetry group $SU(\cN)$, the sGoldstini transform as
 \begin{align}
  {\scriptsize \yng(1)} \otimes {\scriptsize \yng(1)} = {\scriptsize \yng(2)} \oplus {\scriptsize \yng(1,1)} \,.
 \end{align}
The gauge vectors in the gravity multiplet always transform in the latter representation, allowing for the above interpretation. This is consistent with previously considered cases. First of all, in $\cN =1$ there is no anti-symmetric representation, and indeed the introduction of F-terms does not correspond to a gauging. The symmetric representation is one-dimensional, corresponding to the one physical sGoldstino. In $\cN =2$ the anti-symmetric representation is one-dimensional. Indeed it was found in \cite{Louis}, in the case of only hypermultiplets, that this direction in the scalar manifold corresponds to a gauged isometry, as in \eqref{sG-gaugeinv}. Similarly, the no-go theorem for stable De Sitter in that case was derived from the trace over the three sGoldstini masses in the symmetric representation, corresponding to \eqref{sG-symm}. 

In the $\cN = 4$ case at hand, the trace over sGoldstini masses corresponds to the following projection of the full mass matrix. In the case of  sGoldstini in the gravity sector ($\fo=0$), one should consider the $SL(2)$ scalar mass
 \begin{align}
  M^2_{\rm sG} = V_{(ij)}{}^{\bar \tau} \, \left[m^{2}\right]_{\bar \tau}{}^{\tau} \, V_{\tau}^{(ij)} = \tfrac{1}{48} \, \Fm \, ( - \tfrac16 \Fp - \tfrac16 \Fm + \tfrac12 \Fo ) \,.
 \label{sG-grav}
 \end{align}
In the case of matter sGoldstini ($\fm=0$), the relevant combination is
 \begin{align}
  M_{\rm sG}^2 & = V_{(ij)}^{\{am\}} \, [m^2]_{\{am\}}^{\quad \; \{bn\}} \, V_{\{bn\}}^{(ij)} = \bP_{am,\, bn} V_{am,\,bn} \,, \label{sG-mat}
 \end{align}
where we have used the projection $\bP$ based on the symmetric sGoldstini directions:
 \begin{align}
  \bP_{am, \, bn} & \equiv 4 \, V_{(ij) \, \{am\}} V_{\{bn\}}^{(ij)} \notag \\
  & =  \tfrac{1}{4} \,[ 2 \delta_{\alpha \beta} (\delta_{mn} \Fo_{a,\,b} - 2 \Fo_{an,\,bm}) + \, \epsilon_{\alpha \beta} \epsilon_{m n q_1 q_2 q_3 q_4}  \, f_{\alpha \, a q_1 q_2} f_{\beta \, b q_3 q_4}] \,. \label{sG-P}
 \end{align}
Finally, in the general case with both $\fm$ and $\fo$, one should add the above two expressions and include the crossterm
\begin{align}
   V_{(ij)}^{\bar \tau} \, \left[m^2\right]_{\bar \tau}{}^{\{am\}} \, V_{\{am\}}^{(ij)} + \, V_{(ij)}^{\{am\}} \, \left[m^{2}\right]_{\{am\}}{}^{\tau} \, V_{\tau}^{(ij)} \,.
 \label{sG-cross}
 \end{align}
In the next subsections we will calculate \eqref{sG-grav} and \eqref{sG-mat} explicitly. The general case, including \eqref{sG-cross}, will be beyond the scope of this paper.

In addition to the projection $\bP$ based on the symmetric sGoldstino directions, it will also prove useful to define the similar expression for the anti-symmetric sGoldstini:
 \begin{align}
  \bQ_{am,\, bn} \equiv 4 \, V_{[ij] \, \{am\}} V_{\{bn\}}^{[ij]} = \Fo_{am,\, bn} \,.
 \end{align}
As the antisymmetric sGoldstini directions correspond to gauge directions, this projection annihilates the mass matrix:
 \begin{align}
 V_{am,\, bn} \bQ_{am,\, bn} = 0 \,. \label{sG-gaugeinv2}
 \end{align}
Nevertheless, the projection $\bQ$ will be instrumental in the interpretation of the sGoldstino mass.

\subsection{SUSY breaking in the gravity sector}

First consider the case of supersymmetry breaking due to $\fm$, i.e.~in the gravity sector. In this case the sGoldstino mass is given by \eqref{sG-grav}. Upon properly normalising with respect to the length of the sGoldstino directions  we obtain a new quantity $m^{2}_{\rm sG}$. In units of the scalar potential, it reads
 \begin{align}
   \frac{m^2_{\rm sG}}{V} = \frac{2 \Fp + 2 \Fm}{3 \Fp - \Fm} \,.
 \end{align}
Note that turning on $\fm$ lowers the masses, while it raises the scalar potential - therefore it is clear that there will be some point where the masses become unstable. This happens at $\Fm = \tfrac{1}{11} \Fp$. Note that this transition occurs before the scalar potential becomes zero. Therefore the sGoldstino mass rules out metastable Minkowski or De Sitter solutions with supersymmetry breaking in the gravity sector.

As in the case of supersymmetric vacua, whenever we consider just $\fp$, $\fm$, $\ftw$ and $\fth$, we can write the $SO(6,n)$ mass matrix in the following way
\begin{align}
[m^2]_{\{am\}}^{\quad \; \{bn\}} = - \tfrac18 \, \delta_{ab} \left( 3 \, \fp_{\alpha \, mpq} - \fm_{\alpha \, mpq} \right) \left( 3 \, \fp_{\alpha \, npq} - \fm_{\alpha npq} \right) + \tfrac18 \, V_{am,\,   \alpha \, pqc} V_{bn, \, \alpha \, pqc}
\end{align}
with 
\begin{align}
V_{am, \, \alpha \, pqc} \equiv - 4 \, \delta_{m[p} \, f_{\alpha \, a|q]c} + \,  \delta_{ac} \, \fp_{\alpha \, mpq} - 3 \, \delta_{ac} \, \fm_{\alpha \, mpq}
\end{align}
Note that this is again given by a negative and a positive definite term. With only $\fm$, all scalar masses are equal and positive: $m^2 = \tfrac16 \Fm$. However, including $\ftw$ this degeneracy is lifted and the masses are subject to the lower bound $m^2 \geq - \tfrac{1}{48} \Fm$. Therefore one cannot make definite statements on the $SO(6,6)$ scalars in the general case. 

\subsection{SUSY breaking in the matter sector}

Now let us address the other limiting case, i.e.~supersymmetry breaking in the matter sector. In particular, we will set both $\fp$ and $\fm$ equal to zero, and focus on the roles of $\fo,\ftw,\fth$. Interestingly, we will see that not all lessons from $\cN = 1$ carry over to this case.

In this case supersymmetry breaking proceeds completely via the matter sector, i.e.~the vector multiplets. Calculating the sGoldstino mass \eqref{sG-mat} in this case amounts to
 \begin{align}
  M^2_{\rm sG} = \tfrac{1}{16} \, [ & + 2 \, (\Fo_{a, \, b})^{2} + 12 \, (\Fo_{m,\,n})^{2} - 12 \, (\Fo_{am,\,bn})^{2} - 4 \, (\Fo_{mn,\,pq})^{2} \notag \\
  & + 6 \, \Fo_{a,\,b} \Ftw_{a,\,b} + 4 \, \Fo_{am,\,bn} \Ftw_{am,\,bn} +4\, F^{\text{\tiny{(1 3)}}}_{mn,\,ab} \Fo_{am,\,bn} ]\, .
 \label{sG-first}
 \end{align}
The question is how to interpret this combination, and in particular to find out whether it is positive or negative. Following both the $\cN = 1$ and $\cN =2$ discussion, we will try to write this in terms of a sectional curvature.

The Riemann tensor of the $SO(6,n)$ scalar coset is given  by
 \begin{align}
  R_{am,\,bn,\,cp,\,dq} = - \tfrac12 \, \delta_{mn} \delta_{pq} \, (\delta_{ac} \delta_{bd} - \delta_{ad} \delta_{bc}) -\tfrac12 \, \delta_{ab} \delta_{cd} \, ( \delta_{mp} \delta_{nq} - \delta_{mq} \delta_{np}) \,.
 \end{align}
The natural sectional curvatures in this case are constructed with either the symmetric or the anti-symmetric sGoldstino directions, or both. The corresponding projections are provided by $\bP$ or $\bQ$. This naturally leads to the following three possibilities:
 \begin{align}
  R_{am,\, bn,\, cp,\, dq} \, \bP_{am,\, cp} \, \bP_{bn,\, dq} =& - \tfrac12 \, (\Fo)^{2} + \tfrac32 \, (\Fo_{m,\,n})^{2} - (\Fo_{a,\,b})^{2} + \notag \\
  & - 2 \, (\Fo_{am,\,bn})^{2} + \Fo_{am,\,bn} \Fo_{an,\,bm} - \tfrac12 \, (\Fo_{mn,\,pq})^{2} \,, \notag \\
  R_{am,\, bn,\, cp,\, dq} \, \bP_{am,\, cp} \, \bQ_{bn,\, dq} = & - \tfrac14 \, (\Fo)^{2} + \tfrac12 \, (\Fo_{m,\,n})^{2} - \tfrac12 \, (\Fo_{a,\,b})^{2} - (\Fo_{am,\,bn})^{2} \,, \notag \\
  R_{am,\, bn,\, cp,\, dq} \, \bQ_{am,\, cp} \, \bQ_{bn,\, dq} = & - \tfrac12 \, (\Fo_{m,\,n})^{2} - \tfrac12 \, (\Fo_{a,\,b})^{2} + \Fo_{am,\,bn} \Fo_{an,\,bm}  \,.
 \end{align}
It follows from the geometric properties of the manifold that all sectional curvatures are negative, see e.g.~\cite{Helgason}.

Furthermore we can use the quadratic constraints \eqref{QC}, to which the embedding tensor components are subjected, to try to eliminate the tensors $\ftw$ and $\fth$ from \eqref{sG-first}. Indeed this works for a number of terms appearing in the sGoldstino mass. In particular, we have been able to derive the following relevant quartic relations from \eqref{QC}:
 \begin{align}
 & (\Fo_{mn,\, pq})^2 = 2 \, (\Fo_{m,\,n})^2 + (F^{\text{\tiny(1 2)}}_{mn,\, ap})^2 + 2  \, \Fo_{am,\, bn} F^{\text{\tiny{(1 3)}}}_{mn,\, ab} \,, \notag \\
 & (\Fo_{am,\,bn})^2 = (\Fo_{m,\,n})^2 + \Fo_{am,\,bn} \Ftw_{am,\, bn} \,, \notag \\
 & 2 \, \Fo_{am,\,bn} \Fo_{an,\,bm} = (\Fo_{a,\,b})^2 \,, \notag \\
 & 2 \, \Fo_{am,\, bn} \Ftw_{an,\, bm} = \Fo_{a,\,b} \Ftw_{a,\,b} \,, \notag \\
 & 0 = (- \Fo_{am, \,bn} + \Fo_{an,\, bm} + \Ftw_{am,\, bn} - \Ftw_{an,\,bm} - F^{\text{\tiny{(1 3)}}}_{mn,\, ab}) \, \Fo_{am,\,bn} \,, \label{Quartic relations}
\end{align}
which will be useful in what follows. We have been unable to construct more relations of this form that can be used to massage \eqref{sG-first} into a more managable form.

Using both these quartic relations and the scalar field equations \eqref{N=4-fieldeqs}, the total sGoldstini mass can be written in terms of a single sectional curvature modulo terms proportional to $\Fo \Ftw$ only in the case of $R \, \bQ \, \bQ$:
 \begin{align}
  M^2_{\rm sG} = \tfrac12 \, R \, \bQ \, \bQ + \tfrac12 \, \Fo_{a,\,b} \Ftw_{a,\,b} - \tfrac12 \, \Fo_{am,\,bn} \Ftw_{am,\,bn} - \tfrac14 \, \Fo_{\alpha a, \, \beta b} \Ftw_{\alpha a, \, \beta b} \,.
 \end{align}
The sectional curvature of the coset manifold is negative. Therefore, in the absence $\ftw$, or rather tensors $\Fo$ and $\Ftw$ that can be contracted in non-trivial ways, the sGoldstino mass is always negative. The remaining three terms can be positive, however.

Two features of this sGoldstino mass are surprising from an $\cN =1$ and $\cN =2$ point of view \cite{GR1, Louis}. Firstly,  for $\cN = 1$ and $\cN = 2$ the relevant sectional curvatures are instead $R \, \bP \, \bP$ and $R \, \bP \, \bQ$, respectively. Moreover, we find an explicit appearence of the tensor $\ftw$, associated with supersymmetric mass terms for the vector multiplets\footnote{We have checked explicitly that the Goldstini masses are independent of $\ftw$ and indeed vanish.}. In both the $\cN =1$ and $\cN =2$ cases the corresponding tensor could always be eliminated, at the cost of introducing the supersymmetric AdS scale. In contrast, in $\cN =4$ we have not been able to do so. It cannot be excluded that we have missed a relevant quartic relation in (\ref{Quartic relations}) which would allows one to write the sGoldstino mass purely in terms of the order parameter of supersymmetry breaking. However we do not deem this very likely in view of the following arguments.

The components of $\ftw$ that are picked out by the symmetric sGoldstino directions are not the same components as those constrained by the field equations; these are the two contractions
 \begin{align}
  F^{\text{\tiny(1 2)}}_{mn,\,ap} \, [G^{mnp}]^{ij} \,, \quad
  F^{\text{\tiny(12)}}_{m,\,a} \, [G^m]^{ij} \,, \label{f2-component}
 \end{align}
respectively. This distinction is not present for the cases that have been considered with lower number of supersymmetry, for which these two expressions coincided. In the latter case one can always use the field equation to relate terms like the first in \eqref{f2-component} to the scale of supersymmetric AdS. The fact that this is not possible for a generic $\cN = 4$ configuration can be seen as an explanation for the appearance of the supersymmetric mass term in the sGoldstino mass. 

A second point is the presence of the $\cN = 4$ quadratic constraints. Importantly, these are not necessarily (bi-)linear in the order parameters of supersymmetry breaking, in this case $\fo$; for example, there are relations of the schematic form $( \fo )^2 = (\ftw)^2 $. These do not necessarily hold in the supersymmetric limit, i.e. sending $\fo$ to zero (in contrast to the field equations, which are automatically satisfied in this limit). This has the important implication that for $\cN = 4$ one cannot continuously deform a non-supersymmetric critical point with $\fo \neq 0$ to a supersymmetric critical point with $\fo = 0$. If there would be such a limit, one can argue that the sGoldstini masses must be independent of $\ftw$, as the same holds for the Goldstini masses and these must coincide in supersymmetric limit. In contrast, the absence of such a limit allows the sGoldstini masses to depend on $\ftw$, as indeed we find.

Finally, note that in this case the $SL(2)$ scalars have a positive mass, given by $\tfrac12 \Fp$. However, in this case the $SL(2) \times SO(6,n)$ crossterms in (\ref{mixed-mass-2}) do not necessarily vanish, and hence the mass eigenstates will in general be a mixture of the $SL(2)$ and $SO(6,n)$ scalars.

\subsection{Comparison to literature}

A comparison to the work of \cite{Westra1} is in order at this point. In this work, all semi-simple gauge groups leading to critical points were derived for the special case of six vector multiplets, i.e.~$n=6$. As mentioned before, all De Sitter solutions discussed in that work have only $\fm$ or $\fo$, but not both. They subsequently calculated the scalar mass matrices. In all cases there were no crossterms between the $SL(2)$ and the $SO(6,6)$ scalars, i.e.~(\ref{mixed-mass-2}) vanishes. Furthermore, either the gravity or the matter scalars contain at least one tachyon. We have checked that indeed the gaugings with $\fm$ have an $SL(2)$ instability, while those with $\fo$ are unstable due to $SO(6,6)$ directions. This confirms that the sGoldstini point in approximately the right direction, i.e.~are related to the unstable directions in the models of \cite{Westra1}. However, the sGoldstini are not necessarily mass eigenvalues, and hence are not necessarily identical to the tachyonic directions. Indeed we have seen in a number of cases that the lowest mass satisfies but does not saturate the upper bound set by the sGoldstino mass. In such cases the sGoldstini overlap with the tachyon to a large extent but do not coincide with it (i.e.~their normalised inner product is close to but not equal to one).

The $\cN = 4$ gaugings considered in \cite{RR} are more general in that they have all three tensors $\fp$, $\fm$ and $\fo$. Furthermore, the mixed second derivatives (\ref{mixed-mass-2}) no longer vanishes. In order to derive general results for such gaugings one would have to go beyond the present analysis and include the contributions \eqref{sG-cross} due to mixed supersymmetry breaking.

\section{Discussion and conclusions}

Projecting onto the sGoldstino directions, we have derived an upper bound for the lowest scalar mass of non-supersymmetric critical points of $\cN = 4$ gauged supergravity in a number of cases. The clearest is that of supersymmetry breaking in the gravity sector: the $SL(2)$ scalars in that case always are unstable for Minkowski and De Sitter solutions. The expression in the other case that we have considered, supersymmetry breaking in the matter sector with structure constants $\fo ,\ftw$ and $\fth$, involves a sectional curvature in the $SO(6,n)$ directions. In contrast to the $\cN = 1$ and $\cN =2$ cases, in this case one cannot prove an instability in full generality due to the explicit appearance of the supersymmetric mass terms parametrised by $\ftw$. Instead of a failure to derive a no-go theorem, one could also take this as a hint in order to look for metastable De Sitter solutions. To achieve this, one would have to construct a gauging for which the contribution due to $\ftw$ is positive and outweighs the negative contribution due to the sectional curvature term. It would be interesting to investigate what this implies for the gauge group.

It is clear that the current results only represent a first step towards a full characterisation of the mass of the $\cN =4$ sGoldstino scalars. A further extension would address mixed supersymmetry breaking, and include all five irreducible tensors of $f_{\alpha \, MNP}$. This would be the analogon of F-term supersymmetry breaking in $\cN =1$, with the additional complications of supersymmetry breaking in two sectors and the quadratic constraints associated with non-Abelian gaugings taken into account. No general results exist for other theories wich such features. Owing to the stringent restrictions due to $\cN =4$, this could be the simplest context in which all these complications could be taken into account. However, due to the involved expressions and the possibility to rewrite quantities using quartic relations one would probably have to resort to an automated algorithm to be able to address this case. We summarise the role of the different tensors in table 1.

A further step would be include the two irreducible tensors from $\xi_{\alpha M}$ as well, which we would like to argue is analogous to D-term supersymmetry breaking. The reasons for this interpretation are manifold and include: 
 \begin{itemize}
 \item
 The components giving rise to F- and D-terms are separate irreps of the isometries of the K\"{a}hler manifold, like \eqref{emb-tensor}.
 \item
  The F- and D-term contributions to the scalar potential are indefinite and positive definite, respectively, like \eqref{SW-scalarpot}.
\item
  F-terms give rise to physical sGoldstini, while D-terms also break supersymmetry and only lead to gauge sGoldstini, like \eqref{sGolds}.
 \end{itemize}
However, the $\cN =1$ result in the case with F- and D-terms is strongly model-dependent. Hence it might not expect to be able to derive a general result in a scenario including $\xi$. In particular, as the D-terms contribute to the stability rather than the instability of the sGoldstino scalar, it would appear hard to derive no-go results for $\xi$. Of course, again one can turn this argument around and use the general expressions to look for gaugings that give rise to stable De Sitter critical points. Achieving a positive mass for the sGoldstino directions could be a fruitful guideline when looking for fully stable configurations.

\begin{table}[t]
\begin{center}
\begin{tabular}{||c|c||c|c|c||l||}
\hline 
 & & $\cN = 1$ & $\cN = 4$ & $\cN = 8$ & SUSY ... \\ \hline \hline
$A_1$ & $\bar \psi \psi$ & $\cL$ & $\fp$ & $\bf 36$ & ... AdS \\ \hline
$A_2$ & $\bar \psi \lambda$ & $\cL_i$ & $\fm,\fo$ & $\bf 420$ & ...  breaking \\ \hline
$A_3$ & $\bar \lambda \lambda$ & $\cL_{ij}$ & $\fm, \fo,\ftw$ & $\bf 420$ & ... mass \\ \hline
\end{tabular}
\end{center}
\caption{A comparison of the role of the different tensors in $\cN = 1,4,8$. Only tensors related to F-term scalar potentials are included.} \label{tab:tensors}
\end{table}

Finally, this story does not stop at $\cN = 4$: it is natural to wonder to what extent this method could be used to make definite statements about the case of maximal supergravity. In the $\cN = 8$ case one has 64 complex sGoldstino directions. In terms of the R-symmetry group $SU(8)$, these split up in a $\bf 36$ of symmetric directions, and a $\bf 28$ of anti-symmetric directions. As explained earlier, the latter representation coincides with the gauge vectors, and hence are likely to again correspond to gauge directions. Therefore the symmetric sGoldstini are the physical ones, and should be used to derive lower bounds. Furthermore, in the $\cN = 8$ case there is only the gravity multiplet. Hence one looses the complications due to $\fo$, $\ftw$ and $\fth$. The only tensors correspond to the decomposition of the $\bf 912$ of $E_{7(7)}$ under the R-symmetry group $SU(8)$ \cite{Samtleben}. This gives rise to a complex $\bf 36$, setting the supersymmetric AdS scale, and a complex $\bf 420$, being the order parameter of supersymmetry breaking. Therefore the most general case corresponds to the scenario of supersymmetry breaking in the gravity sector, which was very easy to analyse for the $\cN =4$ case. Of course, the complication one faces at the $\cN =8$ side is that the scalars in the gravity sector span an $E_{7(7)}$ coset, instead of $SL(2)$. Finally, the possibility to add D-terms is absent in this case (as is the overall $U(1)$ factor in the R-symmetry group). Therefore one might hope to be able to derive a fully general no-go theorem for stable and non-supersymmetric Minkowski and/or De Sitter solutions in this case. Note that such a result is not possible for Anti-de Sitter; firstly this seems to be the direction that the interpretation of \eqref{N=1-sG} is heading to, and secondly because of the explicit counterexample of \cite{Warnernew2}.

In addition to metastable supersymmetry breaking in supergravity,  the sGoldstino approach can also be used in globally supersymmetric theories. This was pioneered in \cite{Jacot} for $\cN =1$ and $\cN = 2$. In fact, the present results can be used for the $\cN =4$ globally supersymmetric case after a limiting procedure \cite{Hohm}, in which one eliminates the gravity multiplet and hence the local nature of supersymmetry. Due to the absence of this multiplet, the tensors $\fp$ and $\fm$ drop out in such a limit and the most general analysis is that of section 5.3. It would be interesting to pursue this is more detail, and see whether $\cN = 4$ gauge theories have non-supersymmetric and stable vacua.

\section*{Acknowledgements}

We would like to thank Giuseppe Dibitetto, Adolfo Guarino, Tom\'{a}s Ort\'{\i}n, Jan Rosseel, Jorge Russo, Paul Townsend and the JHEP referee for very useful and interesting discussions. We also thank Jan Rosseel for a careful reading of the manuscript. Furthermore, D.R.~would like to express his gratitude to the LMU M\"{u}nchen for its warm hospitality while part of this project was done. The work of both authors is supported by a VIDI grant from the Netherlands Organisation for Scientific Research (NWO).

\appendix

\section{Conventions}

\subsection{Notation}

Our use of indices is as follows: 
 \begin{align}
  SO(6,n): & \quad M,N,P, \ldots, \notag \\
  SO(6): & \quad m,n,p, \ldots, \notag  \\
  SO(n): & \quad a,b,c, \ldots, \notag \\
  SL(2): & \quad \alpha, \beta, \gamma, \ldots.
 \end{align}
Furthermore, we use the following invariant tensors. The $SO(6,n)$ metric is $\eta_{MN} = (-1,\ldots,-1,+1,\ldots,+1)$. The metrics for the $SO(6)$ and $SO(n)$ parts are both plus one, and hence we do not distinguish between upper and lower indices. The $SO(6)$ invariant Levi-Civita symbol is $\epsilon_{mnpqrs}$ with $\epsilon_{123456} = +1$. In contrast, $SL(2)$ indices are raised and lowered with the invariant Levi-Civita tensor $\epsilon_{\alpha \beta}$ with $\epsilon_{12} = \epsilon^{12} = + 1$.
  
\subsection{$F$-tensors}

We define a tensor $F^{(i)}$ ($F^{(i \, j)}$) taking a suitable contraction of two equal (different) irreducible representations $f^{(i)}$ inside the structure constants.
A comma divides the indices which sit on the first $f$ from those which sit on the second one, e.g.
\begin{align}
\Fo_{m, \, n} = f_{\alpha \, m pc} \, f_{\alpha \, n pc} \, , \quad F^{\text{\tiny{(1 2)}}}_{m, \, a} = f_{\alpha \, m pc} \, f_{\alpha \, a pc} \, .
\end{align}
All indices which are not explicitly given are summed over, where the contractions are performed with $\delta$'s. An important point to notice is that this way of writing is unique.

Due to the fact that $\fp$ is imaginary self-dual, the tensors $\Fp$ has some special features. In particular we have that
\begin{align}
\Fp_{\alpha, \, \beta} & = \tfrac12 \, \delta_{\alpha \beta} \, \Fp \, ,\qquad
\Fp_{m, \, n} = \tfrac16 \, \delta_{mn} \, \Fp \, . \label{RelationsFpm-1}
\end{align}
The same properties are shared by the tensor $\Fm$. Furthermore, the cross terms satisfy
 \begin{align}
 F^{\text{\tiny{(+ --)}}} = 0 \, , \qquad F^{\text{\tiny{(+ --)}}}_{[\alpha, \, \beta]} = 0 \, , \qquad F^{\text{\tiny{(+ --)}}}_{[m, \, n]} = 0 \, . \label{RelationsFpm-2}
 \end{align}

\subsection{Relation between $SU(4)$ and $SO(6)$}

For every pair of anti-symmetric $SU(4)$ indices $[ij]$, we define
\begin{align}
\label{phi down} \phi_{ij} = \tfrac{1}{2} \sum_{m=1}^{6} \phi_{m} \, [G_{m}]_{ij} \, , \qquad
\phi^{ij} & = - \tfrac{1}{2} \sum_{m=1}^{6} \phi_{m} \, [G_{m}]^{ij} \, ,
\end{align}
where the $G$'s are the 't Hooft symbols
\begin{eqnarray*}
[G_{1}]_{ij} = \begin{matr}{cccc} 0 & i & 0 & 0 \\ -i & 0 & 0 & 0 \\ 0 & 0 & 0 & -i \\ 0 & 0 & i & 0 \end{matr} & , & [G_{2}]_{ij} = \begin{matr}{cccc} 0 & 0 & i & 0 \\ 0 & 0 & 0 & i \\ -i & 0 & 0 & 0 \\ 0 & -i & 0 & 0 \end{matr} \, ,\\
\left[ G_{3} \right]_{ij} = \begin{matr}{cccc} 0 & 0 & 0 & i \\ 0 & 0 & -i & 0 \\ 0 & i & 0 & 0 \\ -i & 0 & 0 & 0 \end{matr} & , & [G_{4}]_{ij} = \begin{matr}{cccc} 0 & -1 & 0 & 0 \\ 1 & 0 & 0 & 0 \\ 0 & 0 & 0 & -1 \\ 0 & 0 & 1 & 0 \end{matr} \, ,\\
\left[ G_{5} \right]_{ij} = \begin{matr}{cccc} 0 & 0 & -1 & 0 \\ 0 & 0 & 0 & 1 \\ 1 & 0 & 0 & 0 \\ 0 & -1 & 0 & 0 \end{matr} & , & [G_{6}]_{ij} = \begin{matr}{cccc} 0 & 0 & 0 & -1 \\ 0 & 0 & -1 & 0 \\ 0 & 1 & 0 & 0 \\ 1 & 0 & 0 & 0 \end{matr} \, .
\end{eqnarray*}
For every $m = 1, \, \ldots, \, 6$ we have
\begin{align}
[G_{m}]^{ij} = - \tfrac{1}{2} \epsilon^{ijkl} \, [G_{m}]_{kl} = - ([G_{m}]_{ij})^{*} \, .
\end{align} 
Furthermore, they satisfy the following relations
\begin{align}
[G_m]_{ik} \, [G_{n}]^{kj} + \, [G_n]_{ik} \, [G_{m}]^{kj} & = 2 \, \delta_i^j \, \delta_{mn}  \,, \notag \\
[G_m]_{ik_1} \, [G_{n}]^{k_1k_2} \, [G_p]_{k_2k_3} \, [G_{q}]^{k_3k_4} \, [G_r]_{k_4k_5} \, [G_{s}]^{k_5j} & = -i \, \delta_i^j \, \epsilon_{mnpqrs} \, .
\end{align}
Using these symbols we can construct the Gamma matrices for the eight-dimensional spinorial representation of $SO(6)$:
\begin{equation}
\Gamma_{m} = \begin{matr}{cc} 0 & \left[ G_{m} \right]_{ij} \\ \left[ G_{m} \right]^{ij} & 0 \end{matr} \, .
\end{equation}
Thanks to the properties of the 't Hooft symbols these gamma matrices satisfy the standard Clifford algebra $\Gamma_{(m} \Gamma_{n)} = \delta_{mn} \eins_{8\times8}$ with metric $(+ \, \ldots \, +)$. 

\subsection{$SL(2)/SO(2)$ coset space}

\paragraph{Coset representative in \emph{triangular gauge}.}
The standard way of parametrizing the $SL(2)/SO(2)$ scalar coset is using the triangular gauge. In such a gauge we can write down the coset representative as
\begin{align}
\cV \equiv \exp \left\{ \chi \begin{matr}{cc} 0 & 1 \\ 0 & 0 \end{matr} \right\} \; \exp \left\{ \phi \begin{matr}{cc} -\nicefrac{1}{2} & 0 \\ 0 & \nicefrac{1}{2} \end{matr} \right\} \, ,
\end{align}
and this gives the following expression for the vielbein
\begin{align}
\cV = \begin{matr}{cc} e^{-\nicefrac{\phi}{2}} & \chi e^{\nicefrac{\phi}{2}} \\ 0 & e^{\nicefrac{\phi}{2}} \end{matr} \, .
\end{align}
The metric on the scalar manifold is given by
\begin{align} \label{metrictauphi}
\cM = \cV \, \cV^{T} = \tfrac{1}{e^{-\phi}} \begin{matr}{cc} e^{-2 \phi} + \chi^{2} & \chi \\ \chi & 1 \end{matr} \, ,
\end{align}
and the kinetic Lagrangian is
\begin{align}
\nonumber e^{-1} \mathcal{L}_{\textsc{kin sl(2)}}[\chi,\phi] & = - \tfrac{1}{2} \, \text{tr} \left\{ \cV^{-1} \partial_{\mu} \cV \; \mathbb{P} \; \cV^{-1} \partial^{\mu} \cV \right\}  \\
& = \label{axio-dilaton2} -\tfrac{1}{4} \, (e^{2 \phi} \partial_{\mu} \chi \, \partial^{\mu} \chi + \partial_{\mu} \phi \, \partial^{\mu} \phi) \, .
\end{align}
In terms of the field $\tau = \chi + ie^{- \phi}$ we could write the metric in the following way
\begin{align} \label{metric tau}
\cM = \tfrac{1}{\Im\{\tau\}} \, \begin{matr}{cc} |\tau|^{2} & \Re\{\tau\} \\ \Re\{\tau\} & 1 \end{matr} \, .
\end{align}
and the kinetic Lagrangian is given by
\begin{align}
e^{-1} \cL_{\textsc{kin sl(2)}}[\tau] & = - \, \frac{\partial_{\mu} \tau \partial^{\mu} \bar \tau}{4 \Im^{2}\{\tau\}} \,.  
\end{align}

\paragraph{Coset representative in \emph{unitary gauge}.}
Instead of working in triangular gauge in deriving the scalar masses and the other interesting quantities we have chosen a different gauge. We take the following generators for $SL(2)$:
\begin{align}
\left[ t_{\alpha \beta} \right]_{\gamma}{}^{\eta} = \delta_{(\alpha \phantom{1} \beta) \gamma}^{\eta} \!\!\!\!\!\!\!\!\! \epsilon \quad \; ,
\end{align}
and consider the following expression for the vielbein
\begin{align}
\nonumber \cV & =  \exp \left\{ \phi^{\alpha \beta} \left[t_{\alpha \beta} \right] \right\} \phantom{\dfrac{1}{2}} \\ 
\nonumber & =  \exp \left\{ \phi^{11} \begin{matr}{cc} 0 & 0 \\ 1 & 0 \end{matr} + \phi^{22} \begin{matr}{cc} 0 & -1 \\ 0 & 0 \end{matr} + \, (\phi^{12} + \phi^{21}) \begin{matr}{cc} -\nicefrac12 & 0 \\ 0 & \nicefrac12 \end{matr} \right\} \\
\nonumber & =  \exp \bigg\{ \tfrac{1}{2} (\phi^{11} + \phi^{22}) \begin{matr}{cc} 0 & -1 \\ 1 & 0 \end{matr} + \, \tfrac{1}{2} ( \phi^{11} - \phi^{22} ) \begin{matr}{cc} 0 & 1 \\ 1 & 0 \end{matr} + \, \tfrac12 (\phi^{12} + \phi^{21}) \begin{matr}{cc} -1 & 0 \\ 0 & 1 \end{matr} \bigg\} \\
& =  \exp \bigg\{ \underbrace{\Theta \begin{matr}{cc} 0 & -1 \\ 1 & 0 \end{matr}}_{\text{compact part}} + \, \underbrace{\Xi \begin{matr}{cc} 0 & 1 \\ 1 & 0 \end{matr} + \Phi \begin{matr}{cc} -1 & 0 \\ 0 & 1 \end{matr}}_{\text{non-compact part}} \bigg\} \, .
\end{align}
Here we have defined
\begin{align}
\Theta = \tfrac{1}{2} ( \phi^{11} + \phi^{22} ) \phantom{\dfrac{1}{2}} \,, \quad  \Xi = \tfrac{1}{2} ( \phi^{11} - \phi^{22} ) \phantom{\dfrac{1}{2}} \,, \quad \Phi = \tfrac12(\phi^{12} + \phi^{21}) \phantom{\dfrac{1}{2}}  \, .
\end{align}
We now discard the compact part and consider just the non-compact one:
\begin{eqnarray}
\nonumber \cV & = & \exp \bigg\{ \Xi \begin{matr}{cc} 0 & 1 \\ 1 & 0 \end{matr} + \Phi \begin{matr}{cc} -1 & 0 \\ 0 & 1 \end{matr} \bigg\} \\
& = & \begin{matr}{cc} \cosh \Delta - \dfrac{\Phi}{\Delta} \sinh \Delta & \dfrac{\Xi}{\Delta} \sinh \Delta \\ \dfrac{\Xi}{\Delta} \sinh \Delta & \cosh \Delta + \dfrac{\Phi}{\Delta} \sinh \Delta \end{matr} \, ,
\end{eqnarray}
where $\Delta = \sqrt{\Xi^{2} + \Phi^{2}}$. Starting from the expression of the vielbein we can derive the metric as in (\ref{metrictauphi}).

\paragraph{The origin of coset space.}
In this paper we use some notation directly taken from \cite{SW}. A complex vielbein is defined
\begin{align}
\cV_{\alpha} \equiv \begin{matr}{c} \cV_{\uparrow} \\ \cV_{\downarrow} \end{matr} \, ,
\end{align}
and the metric on the $SL(2)/SO(2)$ sector of the scalar manifold is given by
\begin{align}
\cM_{\alpha \beta} = \Re \left\{ \cV_{\alpha} (\cV_{\beta})^{*} \right\} \, .
\end{align}
At the origin of the moduli space, the metric reduces to $\cM_{\alpha \beta} = \delta_{\alpha \beta}$. As the form of the metric is independent on the gauge choice, we can use it to determine the values of the scalars at the origin. In particular we have
\begin{align}
\chi=\phi=0 \, , \quad \tau = i \, , \quad \Phi=\Xi = 0 \, , \quad \cV_{\alpha} = \begin{matr}{c} -i \\ 1 \end{matr} \, , \quad 
\end{align}
while, at linear order in the origin, we have the following relations
\begin{align}
\chi = \phi^{11} - \phi^{22} \,, \quad \phi =  \phi^{12} + \phi^{21} \, .
\end{align}

\subsection{$SO(6,n)/SO(6) \times SO(n)$ scalar coset}
Here again we follow the conventions of \cite{SW}. The generators are given by
\begin{align}
\left[ t_{TU} \right]_{M}{}^{N} = \delta_{[T  \phantom{M} U]M}^{N} \!\!\!\!\!\!\!\!\!\!\!\!\! \eta \quad \quad  \,.
\end{align}
The vielbein is given by
\begin{align}
\cV\equiv \exp \left\{ \phi^{TU} [t_{TU}] \right\} \, ,
\end{align}
where the summation within the exponential must be taken only on the non compact generators (which are associated to the physical degrees of freedom).

From the vielbein we build up the kinetic terms using \cite{SW}
\begin{align}
e^{-1} \cL_{\textsc{kin so(6,$n$)}} = + \tfrac{1}{16} (\partial_{\mu} \cM_{MN}) (\partial^{\mu} \cM^{MN}) \, .
\end{align}
At lowest order these can be written as
\begin{align}
e^{-1} \cL_{\textsc{kin so(6,$n$)}} = - \tfrac12 \sum_{\{am\}} \partial_{\mu} \phi^{\{am\}} \partial^{\mu} \phi^{\{am\}} \,,
\end{align}
where $\phi^{\{am\}} = \tfrac12 ( \phi^{am} - \phi^{ma} )$.
Furthermore, in the scalar potential we have used the definition
 \begin{align}
  \cM_{M_1 \cdots M_6} = \cV_{M_1}{}^{m_1} \cdots \cV_{M_6}{}^{m_6} \epsilon_{m_1 \cdots m_6} \,.
 \end{align}

 \section{Anti-symmetric sGoldstini as gauge directions}
 In this section we prove (\ref{sG-gaugeinv}) in the general case. Starting from (\ref{mixed-mass-2}), (\ref{SO66-mass-2}) and (\ref{sGolds}), clearly the proof reduces to two statements 
 \begin{align}
 & \sigma_{3}^{\alpha \beta} V_{\alpha \beta, \, bn} f_{\gamma b n q} = \sigma_{1}^{\alpha \beta} V_{\alpha \beta, \, b n} f_{\gamma bn q} = 0 \, , \label{offdiaggaugeinv} \\
 & V_{am, \, bn} f_{\gamma \, bnq} = 0 \, . \label{diaggaugeinv}
\end{align} 
The $q$ index is paired up with a 't~Hooft symbol $[G_{q}]^{ij}$ which tells us we are considering all six anti-symmetric sGoldstino directions to annihilate the squared mass matrix. 

Let's start from (\ref{offdiaggaugeinv}). The $\sigma_1$ and the $\sigma_3$ calculations are completely similar thus we can consider just
\begin{align*}
\sigma^{3}_{\alpha \beta} V_{\alpha \beta ,\, bn} f_{\gamma \, bnq} & = \tfrac12 \sigma^{3}_{\alpha \beta} \epsilon_{\eta \alpha} \, ( F_{\eta n ,\, \beta b}^{\text{\tiny{(1 2)}}} + F_{\beta n,\, \eta b}^{\text{\tiny{(1 2)}}} ) \, f_{\gamma \, bnq} \\
& = \tfrac{i}{2} \sigma^{3}_{\alpha \beta} \sigma^{2}_{\eta \alpha} \, ( F_{\eta n ,\, \beta b}^{\text{\tiny{(1 2)}}} + F_{\beta n,\, \eta b}^{\text{\tiny{(1 2)}}} ) \, f_{\gamma \, bnq} \\
& = - \tfrac12 \sigma^{1}_{\eta \beta} \, 2 f_{\eta \, n p_1 c_1} f_{\beta \, b p_1 c_1} f_{\gamma \, b n q} \, .
\end{align*}
Using the quadratic constraints and the anti-symmetry of $f$, we can trade the last two factors for
\begin{align*}
2 f_{\beta \, b p_1 c_1} f_{\gamma \, b n q} = 2 f_{\beta \, p_2 p_1 c_1} f_{\gamma \, p_2 n q} - f_{\beta \, b p_1 n} f_{\gamma \, b c_1 q} + f_{\beta \, p_2 p_1 n} f_{\gamma \, p_2 c_1 q} \, .
\end{align*}
The first two terms in the sum give zero because of the different symmetry properties of $\sigma^1$ and $f$. The last term is of the form
\begin{align*}
- \tfrac12 \sigma^{1}_{\eta \beta} \, f_{\eta \, n p_1 c_1} f_{\beta \, p_2 p_1 n} f_{\gamma \, p_2 c_1 q} \, .
\end{align*} 
Here we need again the quadratic constraints and the anti-symmetry properties of $f$ to get
\begin{align*}
- \tfrac12 \sigma^{1}_{\eta \beta} \, f_{\eta \, n p_1 p_3} f_{\beta \, p_2 p_1 n} f_{\gamma \, p_2 p_3 q} \, .
\end{align*}
Again this term gives zero because of the different symmetry properties of $\sigma^1$ and $f$. Thus we have proven (\ref{sG-gaugeinv}) for the off diagonal terms of the squared mass matrix.

Let's now turn to (\ref{diaggaugeinv}). The expression we obtain has terms which do contain $\epsilon$ tensors explicitly and other ones which do not. We start manipulating the former ones
 \begin{align*}
 - \tfrac12 \epsilon_{\alpha \beta} \epsilon_{m n p_1 p_2 p_3 p_4} f_{\alpha \, a p_1 p_2} f_{\beta \, b p_3 p_4} f_{\gamma \, b n q} \, .
 \end{align*}
 Using the quadratic constraints (\ref{QC}) and the anti-symmetry of the $\epsilon$ tensor we have $f_{\beta \, b p_3 p_4} f_{\gamma \, b n q} = f_{\beta \, r p_3 p_4} f_{\gamma \, r n q}$. After some manipulations, exploiting the symmetry properties of $f^{(\pm)}$, this terms can be written as
 \begin{align}
 - f_{\alpha \, a p_1 p_2} ( \fp_{\alpha \, r p_1 p_2} - \fm_{\alpha\, r p_1 p_2}) f_{\gamma \, mqr} - 2 \, f_{\alpha \, a p_1 p_2} ( \fp_{\alpha \, m p_1 p_3} - \fm_{\alpha \, m p_1 p_3} ) f_{\gamma \, q p_3 p_2} \, . \label{epsterm}
 \end{align}
 We now consider the remaining terms. They are given by
 \begin{align}
& - \Fp_{m,\,n} \, f_{\gamma \, anq} + \Fm_{m,\,n} \, f_{\gamma \, anq} + \Fo_{m,\,n} \, f_{\gamma \, anq} + \Ftw_{a,\,b} \, f_{\gamma \, bmq} - \Fo_{an, \, bm} \, f_{\gamma \,bnq} - \Ftw_{an, \, bm} \, f_{\gamma \, bnq} \nonumber \\ 
& - F^{\text{\tiny{(\!+ 2)}}}_{mn, \, ab} \, f_{\gamma \, bnq} + 3 F^{\text{\tiny{(-- 2)}}}_{mn , \, ab} \, f_{\gamma \, bnq} + F^{\text{\tiny{(1 3)}}}_{mn, \, ab} \, f_{\gamma \, bnq} \, . \label{otherterms}
 \end{align}
 Making heavy use of the quadratic constraints it's possible to manipulate the third and fourth term in the first line into
 \begin{align*}
 \Fo_{m,\,n} \, f_{\gamma \, anq} = & - F^{\text{\tiny{(1 2)}}}_{m,\,b} \, f_{\gamma \, aqb} + \Fo_{an ,\,b m} \, f_{\gamma \, b n q} \\
 & - F^{\text{\tiny{(1 2)}}}_{m p_1,\, a p_2} \, f_{\gamma p_1 p_2 q} + F^{\text{\tiny{(1 2)}}}_{mc_1,\, a c_2} \, f_{\gamma \, c_1 c_2 q} + F^{\text{\tiny{(1 3)}}}_{mn, \, ab} \, f_{\gamma \, bnq} \, ,\\
 \Ftw_{a,\,b} \, f_{\gamma \, bmq} = & + f_{\alpha \, a p_1 p_2} ( \fp_{\alpha \, r p_1 p_2} - \fm_{\alpha \, r p_1 p_2} ) f_{\gamma \, mqr} + \Ftw_{an, \, bm} \, f_{\gamma \, bnq} \\
 & - F^{\text{\tiny{(\!+ 2)}}}_{mn,\, ab} \, f_{\gamma \, bnq} - F^{\text{\tiny{(-- 2)}}}_{mn \, ab} \, f_{\gamma \, bnq} + F^{\text{\tiny{(1 2)}}}_{m p_1,\, a p_2} \, f_{\gamma p_1 p_2 q} - F^{\text{\tiny{(1 2)}}}_{mc_1,\, a c_2} \, f_{\gamma \, c_1 c_2 q} \, .
 \end{align*}
 Substituting these relations in (\ref{otherterms}) and including also the term proportional to the $\epsilon$ tensor (\ref{epsterm}), we obtain
 \begin{align*}
 & - \Fp_{m,\,n} \, f_{\gamma \, anq} + \Fm_{m,\,n} \, f_{\gamma \, anq} - 2 F^{\text{\tiny{(\!+ 2)}}}_{mn, \, ab} \, f_{\gamma \, bnq} + 2 F^{\text{\tiny{(-- 2)}}}_{mn , \, ab} \, f_{\gamma \, bnq} \\
 & + 2 \, f_{\alpha \, a p_1 p_2} ( \fp_{\alpha \, m p_1 p_3} - \fm_{\alpha \, m p_1 p_3} ) f_{\gamma \, q p_2 p_3} - F^{\text{\tiny{(1 2)}}}_{m,\, b} \, f_{\gamma \, aqb} \, .
 \end{align*}
 Now we use the stationarity condition (\ref{N=4-fieldeqs}) and the quadratic constraints to manipulate the last term in the sum
 \begin{align*}
 - F^{\text{\tiny{(1 2)}}}_{m,\, b} \, f_{\gamma \, aqb} & = - (\fp_{\alpha \, p_1 p_2 m} - \fm_{\alpha \, p_1 p_2 m} ) \, f_{\alpha \, p_1 p_2 b} \, f_{\gamma \, aqb} \\
 & = - (\fp_{\alpha \, p_1 p_2 m} - \fm_{\alpha \, p_1 p_2 m} ) \, ( 2 f_{\alpha \, a p_2 b} f_{\gamma \, p_1 q b} - 2 f_{\alpha \, a p_2 p_3} f_{\gamma \, p_1 q p_3} + f_{\alpha \, p_1 p_2 p_3} f_{\gamma \, a q p_3} ) \, .
 \end{align*}
 Substituting back and using (\ref{RelationsFpm-2}) we get the desired zero.

\providecommand{\href}[2]{#2}\begingroup\raggedright\endgroup

\end{document}